# Deep Compressive Macroscopic Fluorescence Lifetime Imaging

*Ruoyang Yao, Marien Ochoa, Xavier Intes and Pingkun Yan*

Department of Biomedical Engineering, Rensselaer Polytechnic Institute, Troy, NY, 12180

**ABSTRACT**

Compressive Macroscopic Fluorescence Lifetime Imaging (MFLI) is a novel technical implementation that enables monitoring multiple molecular interactions in macroscopic scale. Especially, we reported recently on the development of a hyperspectral wide-field time-resolved single-pixel imaging platform that facilitates whole-body *in vivo* lifetime imaging in less than 14 minutes. However, despite efficient data acquisition, the data processing of a Compressed Sensing (CS) based inversion plus lifetime fitting remain very time consuming. Herein, we propose to investigate the potential of deep learning for fast and accurate image formation. More precisely we developed a Convolutional Neural Network (CNN) called Net-FLICS (Network for Fluorescence Lifetime Imaging with Compressive Sensing) that reconstructs both intensity and lifetime images directly from raw CS measurements. Results show that better quality reconstruction can be obtained using Net-FLICS, for both simulation and experimental dataset, with almost negligible time compared to the traditional analytic methods. This first investigation suggests that Net-FLICS may be a powerful tool to enable CS-based lifetime imaging for real-time applications.

## 1. INTRODUCTION

Molecular optical imaging is a central tool in numerous biomedical applications with an ever increasing impact on healthcare. Especially, fluorescence based optical techniques at the macroscopic scales have transformed fields such as preclinical imaging or guided surgery thanks to their high sensitivity, specificity and relatively low cost. Moreover, beyond traditional intensity based contrast, fluorescence imaging can uniquely enable monitoring the micro-environments (temperature, pH, $PO_2$, etc.) or nanoscale interactions (such as Förster Resonance Energy Transfer or FRET) via lifetime sensing [1, 2]. However, efficient implementations of macroscopic lifetime imager that operate over extended spectral channels for increased information content are challenging. Recently, we reported on the design and validation of a compressive wide-field time-resolved single-pixel imaging system with hyperspectral capability that fills this gap [3].

During the past decade, spatial light modulators (SLM) like Digital Mirror Devices (DMD) have led to the development of compressive and single-pixel imaging methodologies [4, 5]. Our instrumental approach leverages the use of two DMDs for illumination and detection structured light capabilities, coupled with a timer-solved spectrophotometer to yield whole-body *in vivo* preclinical imaging data over 256 time gates (0-4ns) and 16 wavelength channels. To obtain spatially resolved quantitative lifetime maps, first, single-pixel measurements at each individual spectral channel are used to reconstruct time-resolved intensity images, frame by frame, through some CS-based inverse solvers. Then, least square methods (LSM) based lifetime fitting is performed on the obtained Temporal Point Spread Function (TPSF) for each pixel of these spatial maps to reveal the lifetime image. However, this data processing workflow can be quite time consuming. Hence, there is great interest to develop computationally fast approaches to enable bed-side applications.

Deep learning, a branch of machine learning techniques that relies on multi-layer Neural Networks (NN), has led revolutions in solving long-standing problems in artificial intelligence community as well as a variety of other fields [6]. One of the commonly used architectures, namely Convolutional Neural Network (CNN), has presented record breaking results in computer vision tasks such as object recognition and classification [7]. Researchers in Medical Imaging field are also making great efforts to apply CNN to solve their problems [8], including but not limited to lesion detection, image segmentation, resolution augmentation. Few methods have also been proposed for improving image reconstruction. Kulkarni et al. [9] developed a CNN based ReconNet for reconstructing images from simulated compressed sensed data. Wu et al. [10] used multi-layer perceptual network to estimate lifetime values instead fitting from a histogram of photon counts. Very recently, Sinha et al. [11] employed CNN to recover phase objects from propagated intensity diffraction patterns.

In this paper, we propose to address the time consuming data processing issue of Compressive MFLI using deep learning. Our work makes three major contributions. First, a new CNN called Net-FLICS (Network for Fluorescence Lifetime Imaging with Compressive Sensing) is developed, using multiple strategies to provide an end-to-end solution for simultaneously reconstructing both intensity and lifetime images from compressed time-resolved dataset. To the best of our knowledge, this is the first time that a method is demonstrated to accomplish lifetime image reconstruction from CS data without fitting. Second, compared to the previous analytical approach, the developed Net-FLICS is shown to be able to obtain better and more robust reconstruction for both intensity and lifetime images. Last but most importantly, data processing through Net-FLICS only takes negligible computational time compared to that of the traditional workflow, making real-time compressive MFLI more realistic. The rest of the paper is organized as follows. In Section 2, we introduce the design of Net-FLICS, the way of generating simulated data, and details about the training process. In Section 3, we present training results, and compare the reconstruction performance of the proposed approach against the traditional analytical method. Both simulated and experimental data are used to demonstrate the advantage of Net-FLICS. We provide discussions, summarize our work, and introduce the plan for further development in Section 4.

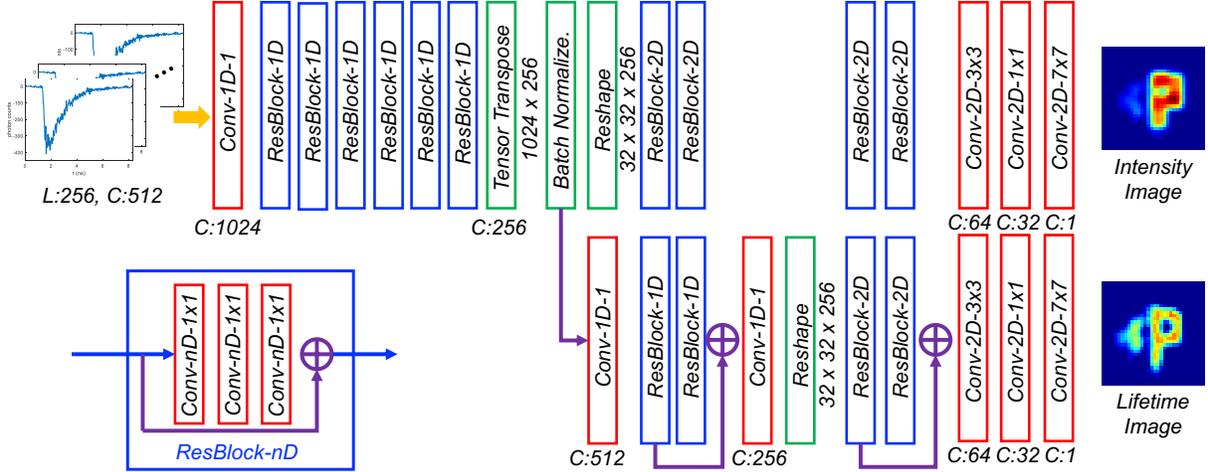

Fig. 1. Overall structure of Net-FLICS.

## 2. DEEP IMAGING

### 2.1. Network Architecture

The overall structure of the proposed Net-FLICS is shown in Fig. 1, where the number and size of the filters used in the convolutional operations are also labeled. Net-FLICS consists of three major segments, a shared convolutional segment, an intensity image reconstruction branch, and a lifetime image reconstruction branch. In our work, the objective is to reconstruct both intensity and lifetime images from the input compressively sensed data. Therefore, a multi-output network is designed, where the first segment of the network is shared by both reconstruction tasks. This shared segment mainly operates on the pattern orientation and aims to recover the sparsity information from the compressively sensed data for later image reconstruction. Once that is achieved, the tensor data are transposed to be processed along the temporal dimension and split to feed into the reconstruction branches. For intensity image reconstruction, the imaging data will be directly reshaped into 32x32x256 for processing and followed by 2D convolutions. For lifetime reconstruction, Net-FLICS will continue with 1D convolutions to mimic the curve fitting process. After that, the data will be reshaped into 32x32x256 for 2D convolutional computation.

The basic building blocks in Net-FLICS are convolutional layers in both 1D and 2D. Each convolutional layer is also followed by a ReLU activation layer. To simplify the structure graph, ReLU layers are not shown in Fig. 1. In addition to the basic convolutional layers, residual connections are heavily used in our work to extend the depth of Net-FLICS. Residual connection is a kind of skip connection that bypasses the non-linear transformations to effectively deal with the gradient vanishing problem in deep neural networks [12]. It has been shown to be a very effective technique for improving the performance of deep networks in a wide range of applications. In our work, we use residual connections not only for constructing ResBlocks but also for connecting ResBlocks in lifetime branch for better reconstruction quality, which is shown to be an effective way for exploiting multi-scale information for reconstruction. At the end of both reconstruction branches, there are three convolutional layers for final image filtering, which are taken from ReconNet [9] due to the proved effectiveness.

### 2.2. Training Data Preparation

In this specific task, the amount of data acquired during imaging experiments is quite limited, since the system has been introduced only very recently [3]. To effectively train the designed Net-FLICS, we created a simulated dataset for training and validation, and the experimental dataset was used later for testing. The MNIST dataset, originally created for handwriting digits recognition [13], is employed to generate intensity and lifetime images and raw imaging data for the network training.

The simulation dataset is created as follows. First, a 28×28 binary image from MNIST is padded with 0s on all sides to reach 32×32 image size. The resulted image has 1s for the digit in it and 0s for background, as shown in Fig. 2(a). To generate an intensity image, we randomly choose several pixels in the foreground and create a 3×3 bright spot around each selected pixel. The intensities of bright spots range from 256 to 1024, and the rest of foreground pixels are randomly assigned intensity values ranging from 64 to 256 (Fig. 2(b)). The lifetime image is similarly created, except that the lifetime values range from 0.3 to 1.5 ns for selected spots and from 0.6 to 1.2 ns for the rest of the object (Fig. 2(c)). Both images are smoothened with a 2D Gaussian filter with kernel size of 1.

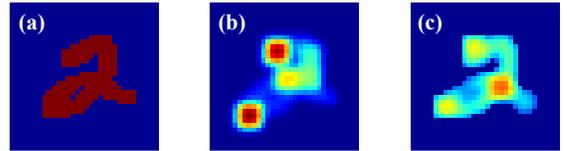

Fig. 2. Compressive MFLI simulation data generation: (a) binary image, (b) intensity image and (c) lifetime image.

Once the intensity and lifetime images are created, we are able to calculate the Temporal Point Spread Function (TPSF) for each pixel with intensity, lifetime, and Instrumental Response Function (IRF) of the system. In this study, we use 512 Hadamard patterns (ranked by spatial frequency) for illumination, and a uniform pattern for detection. Each pattern is divided into two, containing either a positive or a negative component. We sum up the TPSFs for valid pixels in a pattern pair to get two compressed TPSFs. Poisson noise are then added to the integrated curves, and the difference between those curves is used as the final single-pixel measurement (top left in Fig. 1).

## 2.3. Data Augmentation and Optimization

To avoid the potential foreground pixel distribution bias in the training dataset, we actually generate 6 samples for each image in MNIST: the original one, rotating by 90º, rotating by 180º, rotating by 270º, flipping up/down, and flipping left/right. Therefore, in total 60,000 samples are generated from the original MNIST dataset. We used 32,000 samples for training and 8,000 samples for validation. The mean squared error (MSE) between the ground truth and prediction output is defined to be the loss. Since we estimate the two images simultaneously as shown in Fig. 2, there are two loss functions to be minimized, one for intensity image and one for lifetime image. The commonly used RMSprop algorithm is chosen as the optimizer for minimization with a learning rate of 5e-5. The maximal number of epochs for training is set to be 100. We also employed the early stopping strategy with patience of 5. Therefore, the training will stop when either there is no improvement for loss minimization on the validation set for five continuous epochs, or the maximal number of epochs is reached.

## 3. RESULTS

### 3.1. Training Results

The designed Net-FLICS is implemented by using Keras [14], a high-level neural networks API, running on top of TensorFlow framework [15]. With a batch size of 32, the numbers of training and validation samples are 32,000 and 8,000. The training stops after 53 epochs, where each epoch takes 6-7 minutes on a NVIDIA TITAN Xp GPU. The mean absolute error (MAE) values of intensity and lifetime reconstruction during the training process for both dataset are shown in Fig. 3.

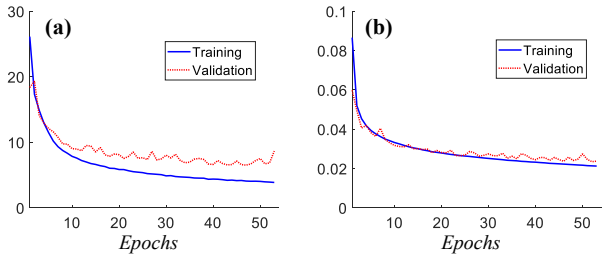

Fig. 3. MAE of intensity (a) and lifetime (b) reconstructions during training.

### 3.2. Performance on Simulated Data

We compare the reconstruction performance of Net-FLICS and the traditional processing workflow on 1000 samples that are not included in either training or validation dataset. For the latter one, we use TVAL3 [16] as the CS inverse solver to recover time-resolved intensity images. Lifetime fitting is performed if the pixel's intensity $I_p$ is larger than 5% of the maximal intensity value $I_{max}$, i.e. $I_p > 0.05 \times I_{max}$. Note that if a TPSF is too noisy for the fitting function to provide robust prediction, the lifetime will be set to the minimum value - 0.3 ns. The histograms of intensity and lifetime MAE for Net-FLICS and TVAL3+ fitting are shown in Fig. 4. As one can see, the average MAE of intensity image reconstruction is only 6.0 for Net-FLICS, compared to 16.2 for TVAL3+fitting, while the average lifetime MAE for Net-FLICS is only 0.025, less than half of 0.057 for TVAL3+fitting. With the proposed deep learning approach, we improve the reconstruction accuracy for more than 60% for both intensity and lifetime images. More importantly, the average time cost for the analytical approach is 35 s, while the average computation time for Net-FLICS is only 4ms for per sample.

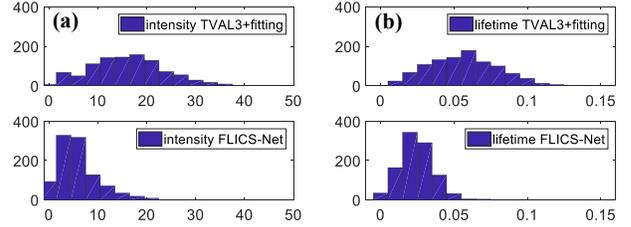

Fig. 4. MAE of intensity (a) and lifetime (b) reconstructions for 1,000 simulated samples.

An example for demonstrating the difference between the two approaches is shown in Fig. 5. Intensity images of ground truth, TVAL3, and Net-FLICS reconstruction are displayed on the top row (left to tight), and corresponding lifetime images are displayed on the bottom. The difference between intensity is not very obvious, but we can notice that values of some pixel are missing for TVAL3 reconstruction. This might be explained by the incompleteness of Hadamard base because we are only using half (512/1024) of the patterns for faster imaging. However, the difference between lifetimes is quite significant, especially for pixels with low intensity. Since the number of photon counts is very small, LSM-based fitting is unable to give reliable prediction due to the presence of strong Poisson noise. This case illustrates the general situation reflected by the statistics in Fig. 4, where both intensity and lifetime reconstructions from Net-FLICS have much better quality than those from TVAL3+fitting.

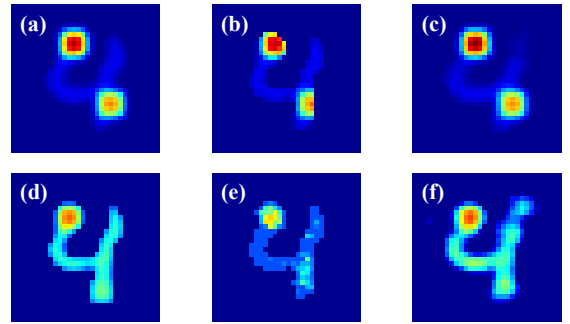

Fig. 5. Intensity (a-c) and lifetime images (d-f) of ground truth (a,d), TVAL3+fitting (b,e), and Net-FLICS (c,f) reconstruction.

### 3.3. Performance on Experimental Data

We also validated the performance of Net-FLICS on experimental MFLI dataset acquired with the wide-field time-resolved single-pixel imaging system. Here, the model trained by using the simulation data is used directly without fine tuning. We prepared a phantom of three letters "RPI" with two dyes, AF750 for 'R' and 'I', and HITCI for 'P'. The excitation wavelength was 740 nm and the laser power was 450 mW. The same 512 Hadamard pattern pairs as above were used for illumination, and the exposure time was 1 second for each pair, leading to a total acquisition time of ~9 minutes. The processing time TVAL+ fitting was 28 seconds in this case. The normalized intensity reconstructions of the ground truth, analytical approach and Net-FLICS are shown in Figs. 6(a-c), and the lifetime results in Figs. 6(d-f).

Although letters 'R' and 'I' are missing from Net-FLICS reconstructions, because their intensities are very low compared to letter 'P', we are able to reconstruct the latter one with high accuracy. In the reconstruction results of Net-FLICS, the intensity is smoother than the result of analytical approach and the average lifetime of 'P', 0.97 ns exactly matches the ground truth value, compared to 0.77 ns from lifetime fitting. Note that the actual patterns generated by DMD contain inhomogeneities and are not perfectly binary, which are different from we used to generate the training data. Thus, there exists certain mismatch in our model. LSM-based lifetime fitting is not optimal either as the values for 'R' and 'I' are also missing due to noisy TPSF curves.

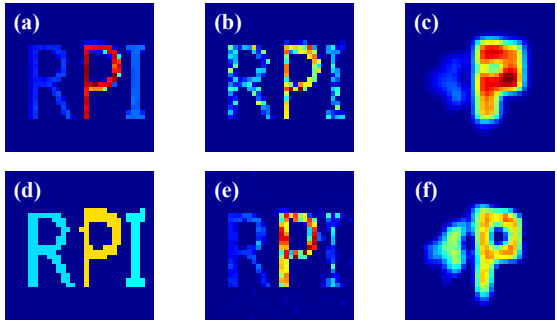

Fig. 6. Intensity (a-c) and lifetime images (d-f) of ground truth (a,d), TVAL3+fitting (b,e) and Net-FLICS (c,f) for experimental dataset.

## 4. CONCLUSION

In conclusion, we have proposed a new CNN called Net-FLICS to reconstruct both intensity and lifetime images directly from compressed time-resolved data from a novel optical imaging system. Compared to the old analytical workflow, Net-FLICS could provide improved results both *in silico* and *in vitro*, and with almost negligible time. The second part of the network can be applied in many other non-CS optical techniques. Net-FLICS could still be improved in several ways. For training data, we will incorporate real CCD patterns, shift TPSF curves to mimic laser jitter, and add more samples besides hand-writing digits to mimic real biological samples. For Net-FLICS itself, we will carefully examine the contribution of each layer and check if we can simplify the structure while preserving its performance. We will also test different parameters such as loss function, learning rate, activation function, etc., to achieve optimal performance. The ultimate goal is to design a upgraded version of Net-FLICS to handle more challenging bi-exponential life fitting, and to process compressed hyperspectral time-resolved data in real-time. The combination of Net-FLICS short computational times and fast imaging acquisition protocols with for instance sparse time information [17] is expected to lead to implementation of wide-field lifetime imaging in time-constrained scenarios such as intraoperative guided surgery or preclinical pharmacokinetics studies. More thorough comparison of the proposed approaches will be carried out in the near future with extensive experimental data sets.

## 5. ACKOWLEDGMENT

This work is supported by the National Institute of Health (NIH) Grants R01 EB19443 and R01 CA207725. We would like to thank NVIDIA for the GPU donation.